\documentclass[twocolumn]{aastex631}

\usepackage{graphicx}
\usepackage{amsmath}
\usepackage{xspace}
\usepackage{tabularx}
\usepackage[hang,flushmargin]{footmisc}

\newcommand{\hst}[0]{\textit{HST}\xspace}
\newcommand{\gaia}[0]{\textit{Gaia}\xspace}

\begin{document}

\title{The internal kinematics of NGC\,2808 and its multiple populations}

\correspondingauthor{M.\ Griggio}
\email{mgriggio@stsci.edu}

\author[0000-0002-5060-1379]{M.\ Griggio}
\affiliation{Space Telescope Science Institute,
3700 San Martin Drive,
Baltimore, MD 21218, USA}

\author[0000-0003-3858-637X]{A.\ Bellini}
\affiliation{Space Telescope Science Institute,
3700 San Martin Drive,
Baltimore, MD 21218, USA}

\author[0000-0003-1820-7985]{F.\ I.\ Aros}
\affiliation{Department of Astrophysics, University of Vienna, Türkenschanzstrasse 17, 1180 Vienna, Austria}

\author[0000-0003-2742-6872]{E.\ Vesperini}
\affiliation{Department of Astronomy, Indiana University, Bloomington, Swain West, 727 E. 3rd Street, IN 47405, USA}

\author[0000-0001-9673-7397]{M.\ Libralato}
\affiliation{INAF - Osservatorio Astronomico di Padova,
Vicolo dell'Osservatorio 5, 35122 Padova, Italy}

\author[0000-0003-2861-3995]{J.\ Anderson}
\affiliation{Space Telescope Science Institute, 3700 San Martin Drive, Baltimore, MD 21218, USA}

\author[0000-0002-1959-6946]{H.\ Baumgardt}
\affiliation{School of Mathematics and Physics, The University of Queensland, QLD 4072, Australia}

\author[0000-0002-2165-8528]{F.\ R.\ Ferraro}
\affiliation{Dipartimento di Fisica e Astronomia, Universit\`a di Bologna, Via Gobetti 93/2 40129 Bologna, Italy}
\affiliation{INAF - Osservatorio di Astrofisica e Scienze dello Spazio di Bologna, Via Gobetti 93/3 40129 Bologna, Italy}

\author[0000-0001-7827-7825]{R.\ P.\ van\ der\ Marel}
\affiliation{Space Telescope Science Institute, 3700 San Martin Drive, Baltimore, MD 21218, USA}
\affiliation{Center for Astrophysical Sciences, The William H. Miller III Department of Physics \& Astronomy, Johns Hopkins University, Baltimore, MD 21218, USA}

\author[0000-0003-4592-1236]{S.\ Raso}
\affiliation{Liceo Marconi-Delpino, 16043 Genova, Italy}

\author[0000-0002-7093-7355]{A.\ Renzini}
\affiliation{INAF - Osservatorio Astronomico di Padova,
Vicolo dell'Osservatorio 5, 35122 Padova, Italy}

\author[0000-0002-4410-5387]{A.\ Rest}
\affiliation{Space Telescope Science Institute,
3700 San Martin Drive,
Baltimore, MD 21218, USA}
\affiliation{Center for Astrophysical Sciences, The William H. Miller III Department of Physics \& Astronomy, Johns Hopkins University, Baltimore, MD 21218, USA}

\author[0000-0003-0894-1588]{R.\ E.\ Ryan\ Jr.}
\affiliation{Space Telescope Science Institute,
3700 San Martin Drive,
Baltimore, MD 21218, USA}



\begin{abstract}
We use new \hst observations coupled with archival data spanning a total {temporal} baseline of 17\,years to study the internal kinematics of the multiple populations in the globular cluster NGC\,2808 from its center out to $\sim 8$ half-light radii ($r_{\rm h}$). We detect different kinematical behaviors between the first- and second-generation populations. This is especially evident towards the external regions of the cluster, where second-generation stars are increasingly more radially anisotropic. Our results are in agreement with theoretical simulations that predict that second-generation stars, initially more concentrated in the inner regions, gradually diffuse outward and develop a stronger radially anisotropic velocity distribution with respect to the first-generation stars. We find the central regions of the cluster to exhibit a higher degree of energy equipartition than the outskirts; our analysis reveals similar levels of energy equipartition in the radial and tangential components of the motion within about 4\,$r_{\rm h}$, while outside 4\,$r_{\rm h}$ the data suggest that the equipartition level of the radial component of the velocity dispersion is slightly higher than that of the tangential component. Finally, we measured the dispersion of the angular momentum $L_z$ for the three main subpopulations along the main sequence, which provides further evidence of the differences in the velocity anisotropy of 1G and 2G stars and shows marginal evidence for the most extreme second-generation subpopulation being slightly more radially anisotropic than the other second-generation subpopulation.
\end{abstract}



\section{Introduction}

The presence of multiple populations (MPs) in globular clusters (GCs) is one of the most astonishing discoveries of the {last} few decades in the field of stellar populations. Once believed to be simple stellar populations -- i.e. composed by stars born in a single burst and with homogeneous chemical composition -- GCs have served as ideal laboratories to test the predictions of stellar evolution models for over half a century \citep[see, e.g.,][]{2006essp.book.....S}.  The detection of MPs in virtually all GCs {\citep[e.g.,][]{2015AJ....149...91P}} has revealed a much more complex picture, raising numerous new questions concerning star formation and dynamical processes governing the formation and evolution of MPs \citep{2004ARA&A..42..385G,2012A&ARv..20...50G,2018ARA&A..56...83B}. The presence of MPs within GCs has been confirmed by several spectroscopic studies, {which} detected variations of light elements such as Na, O, Al, Mg, and He \citep[see, e.g.,][and references therein]{2005ApJ...621..777P,2012A&A...539A..19G,2011A&A...531A..35P,2012ApJ...748...62V,2009A&A...505..139C,2009A&A...505..117C,2010ApJ...722.1373J,2019MNRAS.487.3815M}.  Thanks to the exquisite capabilities of the {\it Hubble Space Telescope} (\hst), it is 
also possible to characterize the MPs over a wide range of masses, both photometrically, using color-magnitude diagrams where MPs appear as distinct multiple sequences if observed with the appropriate filters \citep[see, e.g.,][for a review]{2022Univ....8..359M}, and kinematically, by measuring their motion
\citep[see, e.g.,][]
{2010ApJ...710.1032A,2023ApJ...944...58L,2018ApJ...861...99L,2015ApJ...810L..13B,2018ApJ...853...86B}.
This observational evidence poses several challenges to 
theoretical models of GCs formation and evolution.

While extensive photometric and spectroscopic studies of MPs have been carried out, there remains a notable lack of astrometric analyses of GCs internal kinematics.  
Cluster{s} with different formation and dynamical histories are expected to imprint different kinematic behaviors on their MPs
\citep[see, e.g.,][]{2010ApJ...724L..99B,2022MNRAS.517.1171L,2013MNRAS.429.1913V,2015MNRAS.450.1164H,2016ApJ...823...61M,2019MNRAS.487.5535T,2021MNRAS.502.4290V,2021MNRAS.502.1974S,2024MNRAS.534.2397L,Dalessandro_2024,2025arXiv250502921A}.
Populations tend to dynamically mix more rapidly in the core, so these kinematic imprints may have faded away there, but stars in the outer regions of GCs can still show some signatures of their formation histories.

Here we present a study of the internal kinematics of the GC NGC\,2808.
This cluster, located at a distance of 10.06\,kpc from the Sun \citep{2021MNRAS.505.5957B}, is one of the 
most massive Galactic GCs \citep[$M=7.91\times10^5$\,M$_\odot$,][]{2018MNRAS.478.1520B}, and previous studies \citep[see, e.g.,][]{2007ApJ...661L..53P} have shown that it hosts at least three subpopulations along the main-sequence, a first-generation (1G) population (hereafter referred to as rMS), and two second-generation (2G) populations with enhanced helium content, labeled mMS and bMS. Later, \citet{2015ApJ...808...51M} showed that the {red giant branch (RGB)} can be split into five different populations, and tentatively identified five populations along the cluster main-sequence as well, dividing the rMS into three subpopulations.

From a kinematic standpoint, \cite{2015ApJ...810L..13B} has shown that in intermediate regions of the cluster (between 1.5 and 2\,$r_{\rm h}$, where $r_{\rm h}$ is the half-light radius), the velocity distribution of the 2G populations is radially anisotropic, with a smaller tangential velocity dispersion than the 1G populations.
N-body simulations indicate that the observed differences in the velocity dispersions align well with the expected kinematic signature of the diffusion process, where 2G stars move outward from their initial concentration in the inner regions of the cluster to the outer areas.

To fully understand the possible evolutionary paths of 1G and 2G kinematic properties and how they depend on initial conditions, it is critical to include more distant regions of the cluster.  This is the focus of the present paper.  In fact, the outermost areas are crucial for a complete characterization of a cluster's kinematic properties and for gaining insights into the potential effects of the external Galactic tidal field. These effects are expected to influence 
the development of radial anisotropy in the outermost regions, by preferentially removing from the system stars with highly eccentric orbits and 
potentially making the velocity dispersion more isotropic or even slightly tangentially anisotropic.
\\

The paper is organized as follows.  In Sec.\,\ref{sec:obs} we describe the observations and the data-reduction process used to extract photometry and proper motions.
Section\,\ref{sec:mpop} presents the data analysis, 
selection of MPs
on the color-magnitude diagrams, and the observed velocity-dispersion profiles.
In Sec.\,\ref{sec:th} we discuss our results, and we conclude the paper and give our final remarks in Sec.\,\ref{sec:con}.


\section{Observations and data reduction}
\label{sec:obs}

\begin{figure}
    \centering
    \includegraphics[width=\columnwidth]{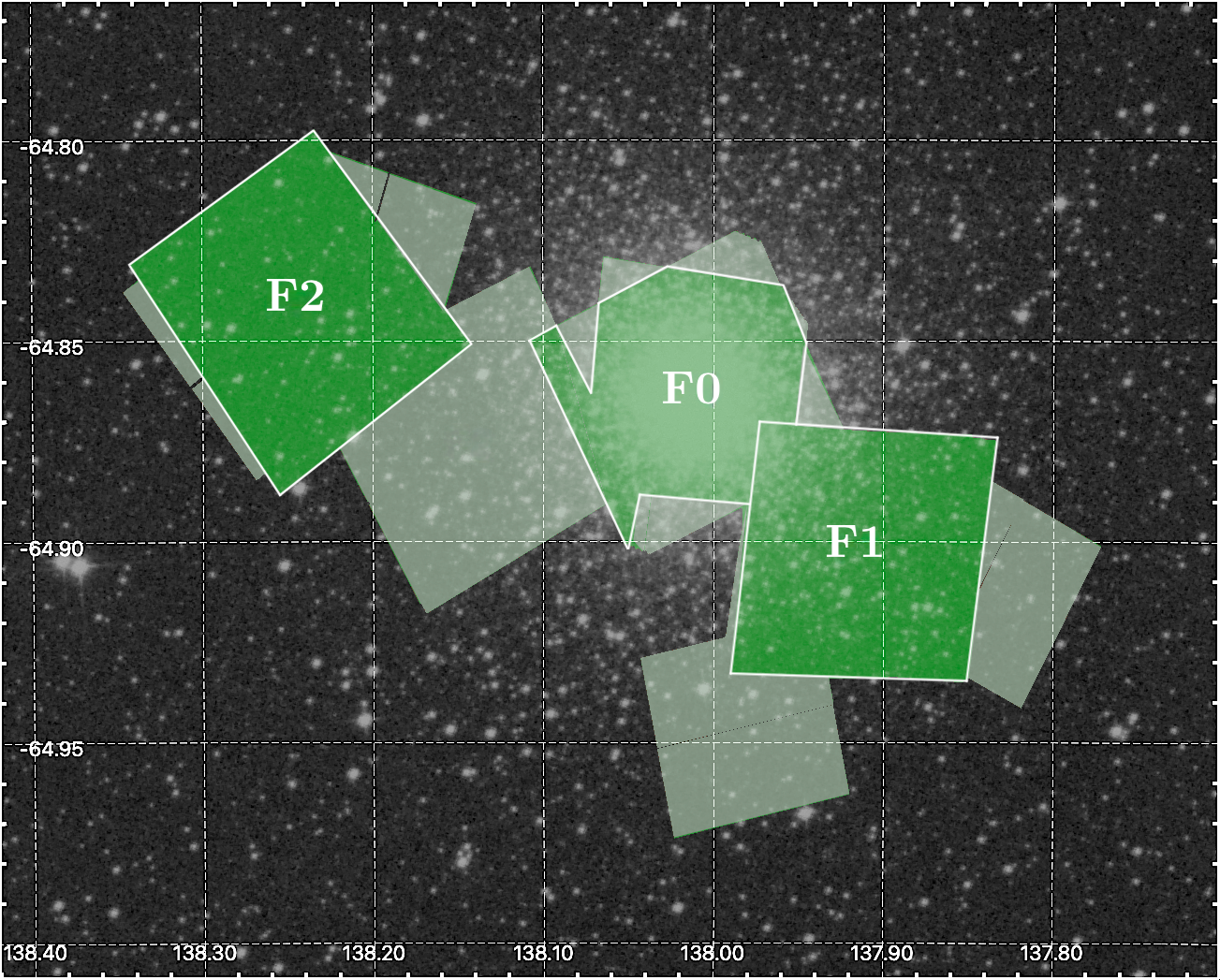}
    \caption{View of NGC\,2808 from the Digitized Sky Survey with the area covered by \hst observations used in this work highlighted in different shades of green. The regions colored in a darker green shade (named F0, F1 and {F2}) are those covered by at least two epochs.}
    \label{fig:fov}
\end{figure}

\begin{table*}
    \centering
    \begin{tabularx}{\textwidth}{lllllll}
    \hline
    \hline
    PI            & GO    & Instr.    & Field & Filter & $N_{\rm images}\times$\,Exposure\,time [s] & Date \\
    \hline
    Piotto        &  9899 & ACS/WFC   & F1    & F475W &  6$\times$340                                                                                        & 2004       \\
    Ford          & 10335 & ACS/HRC   & F0    & F435W & 24$\times$135                                                                                        & 2004--2006 \\
                  &       &           & F0    & F555W &   4$\times$50                                                                                        & 2004--2006 \\
    Ferraro       & 10524 & ACS/WFC   & {F2}    & F555W &  4$\times$250                                                                                        & 2005       \\
    Sarajedini    & 10775 & ACS/WFC   & F0    & F606W &  5$\times$360 $+$ 1$\times$23                                                                        & 2006       \\
                  &       & ACS/WFC   & F0    & F814W &  5$\times$370 $+$ 1$\times$23                                                                        & 2006       \\
    Piotto        & 10922 & ACS/WFC   & F1    & F475W &   1$\times$20 $+$ 2$\times$350 $+$ 2$\times$360                                                      & 2006       \\
                  &       & ACS/WFC   & F1    & F814W &   1$\times$10 $+$ 3$\times$350 $+$ 3$\times$360                                                      & 2006       \\
    Brown         & 11665 & WFC3/UVIS & F1, {F2} & F390W &  12$\times$50 $+$ 6$\times$620 $+$ 6$\times$655 $+$ 1$\times$1020 $+$ 2$\times$1030 $+$ 3$\times$1150& 2010--2011 \\
                  &       & WFC3/UVIS & F1, {F2} & F555W &  6$\times$170 $+$ 3$\times$665 $+$ 3$\times$895                                                      & 2010--2011 \\
                  &       & WFC3/UVIS & F1, {F2} & F814W &   6$\times$50 $+$ 3$\times$1195 $+$ 3$\times$1210                                                    & 2010--2011 \\
    Piotto        & 12605 & ACS/WFC   & {F2}    & F475W &  6$\times$890 $+$ 6$\times$982                                                                       & 2013       \\
                  &       & ACS/WFC   & {F2}    & F814W &  6$\times$508                                                                                        & 2013       \\
                  &       & WFC3/UVIS & F0    & F275W & 12$\times$985                                                                                        & 2013       \\
                  &       & WFC3/UVIS & F0    & F336W &  6$\times$650                                                                                        & 2013       \\
                  &       & WFC3/UVIS & F0    & F438W &   6$\times$97                                                                                        & 2013       \\
    \hline
    Bellini       & 15857 & ACS/WFC   & {F0, F2}    & F475W & 6$\times$819 $+$ 6$\times$878                                                                        & 2020--2021 \\
                  &       & ACS/WFC   & {F0, F2}    & F606W &  5$\times$440 $+$ 4$\times$428                                                                       & 2020--2021 \\
                  &       & ACS/WFC   & {F0, F2}    & F814W &  4$\times$400 $+$ 1$\times$429 $+$ 1$\times$438 $+$ 4$\times$439 $+$ 2$\times$450                    & 2020--2021 \\
                  &       & WFC3/UV   & F1    & F275W & 12$\times$905                                                                                        & 2020--2021 \\
                  &       & WFC3/UV   & F1    & F336W &  4$\times$542 $+$ 2$\times$592                                                                       & 2020--2021 \\
                  &       & WFC3/UV   & F1    & F438W &  2$\times$213 $+$ 4$\times$217                                                                       & 2020--2021 \\
    \hline
    \end{tabularx}
    \caption{Summary of the \textit{HST} data used in this paper. The data are available at\,\dataset[http://dx.doi.org/10.17909/68zd-jx71]{http://dx.doi.org/10.17909/68zd-jx71}.}
    \label{tab:obs_data}
\end{table*}

Our analysis is based on multi-epoch \hst observations of NGC\,2808 that cover the cluster from its core out to $\sim$\,$10\,r_{\rm h}$ \citep[$\sim$\,8\,arcmin, taking $r_{\rm h}=48^{\prime\prime}$,][2010 edition]{1996AJ....112.1487H}. The data used in this paper were collected with the Advanced Camera for Surveys (ACS) and the Wield Field Camera 3 (WFC3) onboard \hst.
As part of a large survey on GCs, we collected new observations of NGC\,2808 in ACS Wide Field Channel (WFC) filters F475W/F606W/F814W and WFC3 UV/Visible channel (UVIS) filters F275W/F336W/F438W (GO-15857, PI: Bellini, see Table\,\ref{tab:obs_data}).
This new dataset, coupled with archival data, yields a total {temporal }baseline of 17 years. A summary of the archival observations used in this study is given in Table\,\ref{tab:obs_data}.

In Fig.\,\ref{fig:fov} we show the area covered by the observations used in this paper. The dark green regions represent the areas covered by at least two epochs (separated by more than 1\,yr) where we can compute precise proper motions.  The fields that are relevant for the present discussion{, i.e., those for which we derived proper motions,} are labeled F0, F1 {and} F2. Field F0 covers the core of NGC\,2808, and it is the field that was used in the previous kinematic study of \cite{2015ApJ...810L..13B}. F1 and {F2} cover two external regions at different radial distances from the cluster center. 

The ACS High Resolution Channel (HRC) archival data (GO-10335, PI: Ford) cover the very center of the cluster in the middle of field F0, and is only used to compute proper motions in the highly-crowded core, where the other instruments are limited by source confusion. Unfortunately, the HRC archival data {were} taken through filters that do not allow us to separate the multiple populations in the color magnitude diagram.

\begin{figure*}
    \centering
    \includegraphics[width=.95\textwidth]{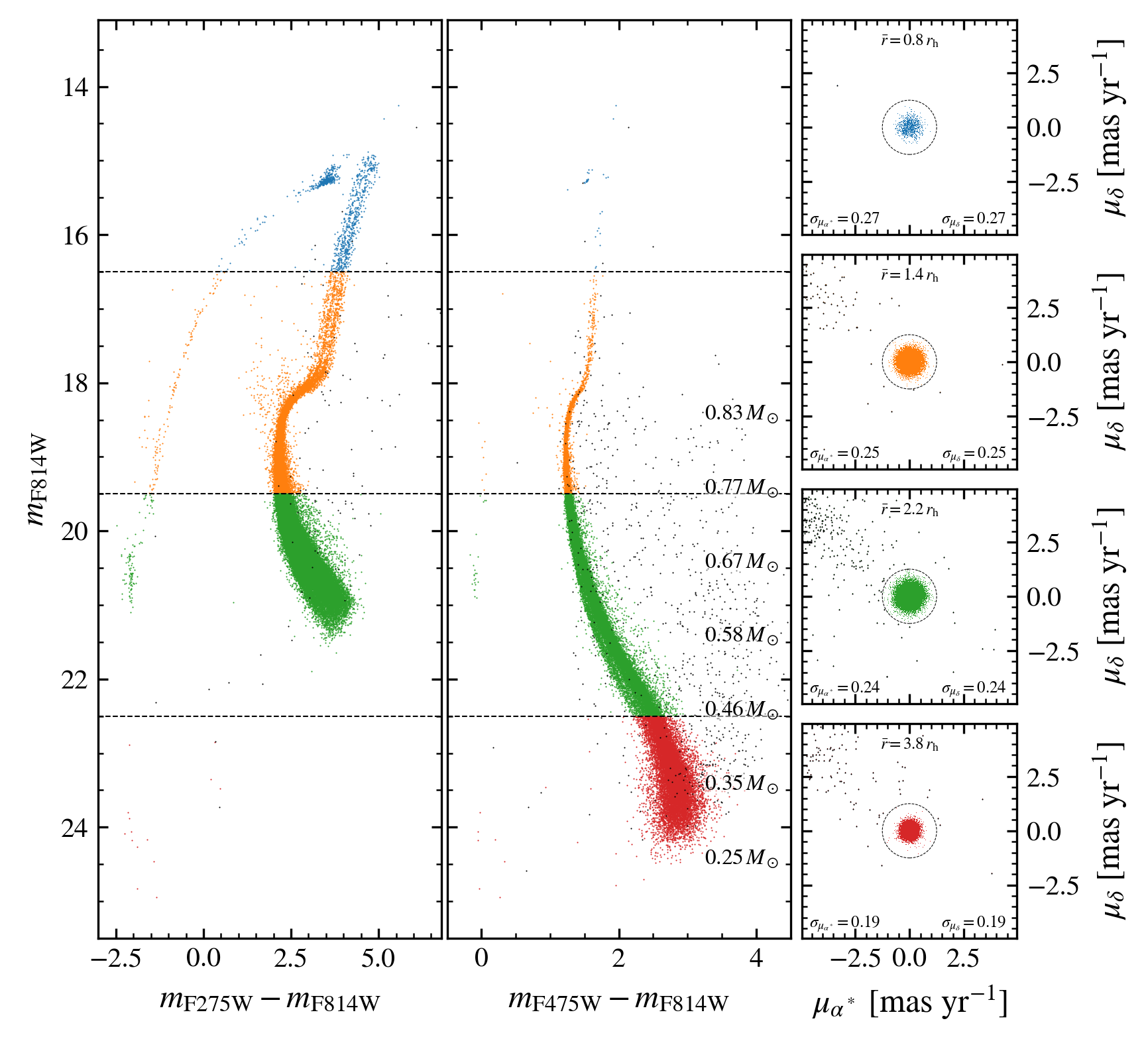}
    \caption{Left: color-magnitude diagram in WFC3/UVIS F275W and ACS/WFC F814W filter{s} {(fields F0 and F1)}. Center: similar to {the} left panel, but with ACS/WFC F475W instead of F275W {(fields F1 and {F2})}. Black dots represent stars with proper motions larger than 1.25\,mas\,yr$^{-1}$ {with respect to the bulk motion of the cluster}. Main-sequence masses for the 1G population are also quoted as a reference of the mass range under study. Masses are based on a solar scaled, 11\,Gyr \citep{2018MNRAS.478.1520B}, [Fe/H]\,$=-1.05$ \citep{2020MNRAS.493.3363H,2024MNRAS.528.1393S} BaSTI isochrone \citep{2018ApJ...856..125H}.
    Right: proper motions for different magnitude intervals. A circle with {a} 1.25\,mas\,yr$^{-1}$ radius is plotted {using} a dashed black line. We report the standard deviation of the {observed} proper-motion distribution along the two axes (for stars inside the circle) in the bottom left and right corners. We also show the average distance from the center, $\bar{r}$, on the top of each panel.}
    \label{fig:cmds}
\end{figure*}

\subsection{First- and second-pass photometry}

The data reduction follows the detailed procedure outlined in, e.g. \cite{2014ApJ...797..115B,2017ApJ...842....6B} and \cite{2018ApJ...861...99L,2022ApJ...934..150L}.  
We used the \texttt{FORTRAN} code \texttt{hst1pass} \citep{2022AAS...24020601A} to find and measure the bright, unsaturated sources in each individual image.  We used pipeline-calibrated, unresampled \texttt{\_flc} frames, which include corrections for detector charge transfer efficiency. 
The code uses focus-diverse (only for ACS/WFC and WFC3/UVIS), library effective point-spread functions \citep[ePSFs][]{2004acs..rept...15A,2018wfc..rept...14A,2018acs..rept....8B}
and perturbs them to account for spatial and temporal variations.
Positions and fluxes measured in each image were corrected for the effects of geometric distortion of the detectors using the distortion solutions provided by \cite{2004acs..rept...15A,2006hstc.conf...11A} for ACS and \cite{2009PASP..121.1419B,2011PASP..123..622B} for UVIS.

The dataset was divided into five groups, combining observations taken within $\sim$\,1\,yr.
Images in each group will be processed together in the next step, allowing us to measure fainter sources. The interval of 1\,yr was chosen such that we have several observations in each group and the {stellar offsets due to} proper motions are mostly negligible within that time span (of the order of  {one} hundredth of an UVIS pixel). 
By leveraging the \gaia \citep{2016A&A...595A...1G} Data Release 3 \citep{2021A&A...649A...1G} absolute reference system \citep[see, e.g.,][]{2023MNRAS.524..108G,2024A&A...687A..94G}, we defined a common reference frame {(the ``master frame'')} for each group by propagating the \gaia positions to the group's mean epoch. We then derived six-parameter linear transformations {(shifts along two directions, change of scale, rotation and skew terms)} to transform the individual-exposure coordinates into the corresponding master frame.

Using the fine-tuned ePSFs and the transformations derived in the previous step, we performed the ``second-pass'' photometry with the \texttt{FORTRAN} code \texttt{KS2} \citep[developed by co-author Jay Anderson, see][]{2017ApJ...842....6B}, considering each group separately.  The \texttt{KS2} code does a series of iterations in which it finds, measures and subtracts sources, addressing progressively fainter objects at each iteration.  The master catalogs created in the previous step are taken as input from the code, and are used to mask bright stars.  The software outputs the positions and fluxes of the stars in the master frame, together with some diagnostic parameters such as the quality of fit (\texttt{qfit}), 
which 
represents the goodness of the ePSF fit to the stars, where $\texttt{qfit}=1$ represents perfect fit of the PSF to the star image, and the \texttt{RADXS} parameter \citep[introduced in][]{2009ApJ...697..965B}, that is indicative of how similar the source flux distribution is to that of a point source.  These parameters can be used to remove bad detections, extended objects, and to reject poorly measured sources.
A detailed description of the \texttt{KS2} code can be found in, e.g., \cite{2017ApJ...842....6B}.
In addition to average positions and magnitudes in the master frame, \texttt{KS2} also provides positions and magnitudes as measured in each individual exposure, which we will use to compute proper motions.

{The photometry was zero-pointed to the Vega-mag system following the prescriptions given in \citep[][]{2017ApJ...842....6B} and using the official aperture corrections and zero points available from the Space Telescope Science Institute (see \citealt{2020acs..rept....8B} for ACS and \citealt{2021wfc..rept....4C} for UVIS).}

\subsection{Proper motions}

Proper motions were measured with the technique introduced by \cite{2014ApJ...797..115B}, and also adopted in, e.g., \cite{2018ApJ...853...86B,2022ApJ...934..150L,2023ApJ...944...58L}.
The process consists of a series of iterations in which we consider each individual image as a standalone epoch, and transform the distortion-corrected positions of each source into the master frame (which is a function of exposure date) by means of a linear relation.  RA and Dec positions are fitted independently. The resulting intercept and slope give the position and motion of the source at the reference time (arbitrarily chosen as the mean epoch), respectively. Propagating these new positions to the average epoch of each group by means of the fitted proper motions allows us to build a new, improved master frame. We then re-derive the transformations from individual exposures into the master frame, using only a local net of cluster members as reference sources.  As a consequence, our measured proper motions are relative to the bulk motion of the cluster. We iterate this process multiple times by refining the master frame and the 
list of reference sources
until the residual between the new positions and those at the previous iteration are smaller than $\sim$\,0.01 pixel.

The measured motions have been corrected for various systematic errors as in \cite{2022ApJ...934..150L}.  To account for low-frequency effects correlated with the temporal baseline and the depth of coverage, we divided the data into groups, according to the available temporal baseline used for calculating the proper motions. For each group, we calculated the median motion along RA and Dec of bona-fide cluster members, i.e. with total proper motions smaller than 1.25\,mas\,yr$^{-1}$, a generous cut since at this point we still have some systematic errors and we do not want to exclude possible members, after having removed outliers.  This value---that should in principle be equal to zero by construction---is then subtracted from each group. High-frequency, spatially and magnitude-dependent systematic errors arise mainly due to charge transfer efficiency {defects} and{,} to a lower extent{,} geometric{-}distortion residuals. To remove residual systematic errors we applied a local correction that, for each target star, is given by the average motion of its surrounding 50 neighbor cluster members within 750 pixels and $\pm$1 $m_{\rm F814W}$.  To select only well measured sources we restricted the sample to stars that have a rejection rate of less than 25 percent, and reduced $\chi^2 < 4$ in each coordinate \citep[see, e.g.][]{2022ApJ...934..150L}. We then iteratively rejected stars with proper-motion errors larger than 50 percent of the local velocity dispersion, as described in \cite{2014ApJ...797..115B}.
\\

Color-magnitude diagrams in the 
two-filter combinations
used in our analysis are shown in Fig.\,\ref{fig:cmds} (left and middle panels). In this figure,  
we only show
the sources that passed our quality cuts in the \texttt{qfit} and \texttt{RADXS} parameters, and we already applied the differential reddening correction \citep[as in][]{2012A&A...540A..16M}.
The $m_{\rm F275W}-m_{\rm F814W}$ diagram covers fields F0 and F1, and the $m_{\rm F475W}-m_{\rm F814W}$ diagram covers F1 and F2, the latter being deeper as it is not limited by the F275W filter.  In the middle panel, on the right side, we report the average mass of 1G stars at different values of $m_{\rm F814W}$.  The 1-mag-bin average masses were estimated using a solar scaled, 11\,Gyr \citep{2018MNRAS.478.1520B}, [Fe/H]\,$=-1.05$ \citep{2020MNRAS.493.3363H,2024MNRAS.528.1393S} BaSTI isochrone \citep{2018ApJ...856..125H}.

Relative proper motions are shown in the right panels of Fig.\,\ref{fig:cmds} for different magnitude intervals.  Black dots in the left and middle panels of Fig.\,\ref{fig:cmds} represent field stars, i.e. those with proper motions larger than 1.25\,mas\,yr$^{-1}$ (about 5 times the observed cluster dispersion).  In each panel, we show the standard deviation along the two axes on the bottom left and right corners, calculated using sources inside the black circle. On top we report the average distance from the cluster center for these stars. These panels are thus not to be interpreted as proper-motion dispersion as function of mass, since stars fainter than $\sim$\,21 in F814W were detected only in the more external regions, where the velocity dispersion is intrinsically smaller. 

\section{The kinematics of the multiple populations}
\label{sec:mpop}
The color-magnitude diagrams we have constructed show that the main sequence of NGC\,2808 splits into at least 3 distinct sequences in the left panels of Fig.\,\ref{fig:sel1} and \ref{fig:sel2}.  They are labeled bMS, mMS and rMS, from the blue to the red.  \cite{2007ApJ...661L..53P} and subsequent papers \citep[see, e.g.,][]{2012ApJ...754L..34M} show that the bMS and the mMS sequences are made of 2G stars, enriched in helium and metals, with $Y\simeq0.38$ and $Y\simeq0.32$ respectively. Using 
pseudo-color
diagrams built by combining \hst filters, \cite{2015ApJ...808...51M} showed that the rMS can itself be divided into three subpopulations with similar He content but with different light element abundances.

However, the separation between the three 1G subpopulations along the MS in the pseudo-color diagram is not so obvious, and it is hard to identify with high confidence samples of stars from each sequence and limit excessive cross contamination. In addition, the separation of the different populations along the RGB is quite challenging in the outer fields due to the available filters and the very small number of evolved stars present (see, e.g., the middle panel of Fig.~\ref{fig:cmds}). For these reasons, in the following analysis we only consider the historical bMS, mMS and rMS populations.

\subsection{Sample selection}

\begin{figure}
    \centering
    \includegraphics[width=\columnwidth]{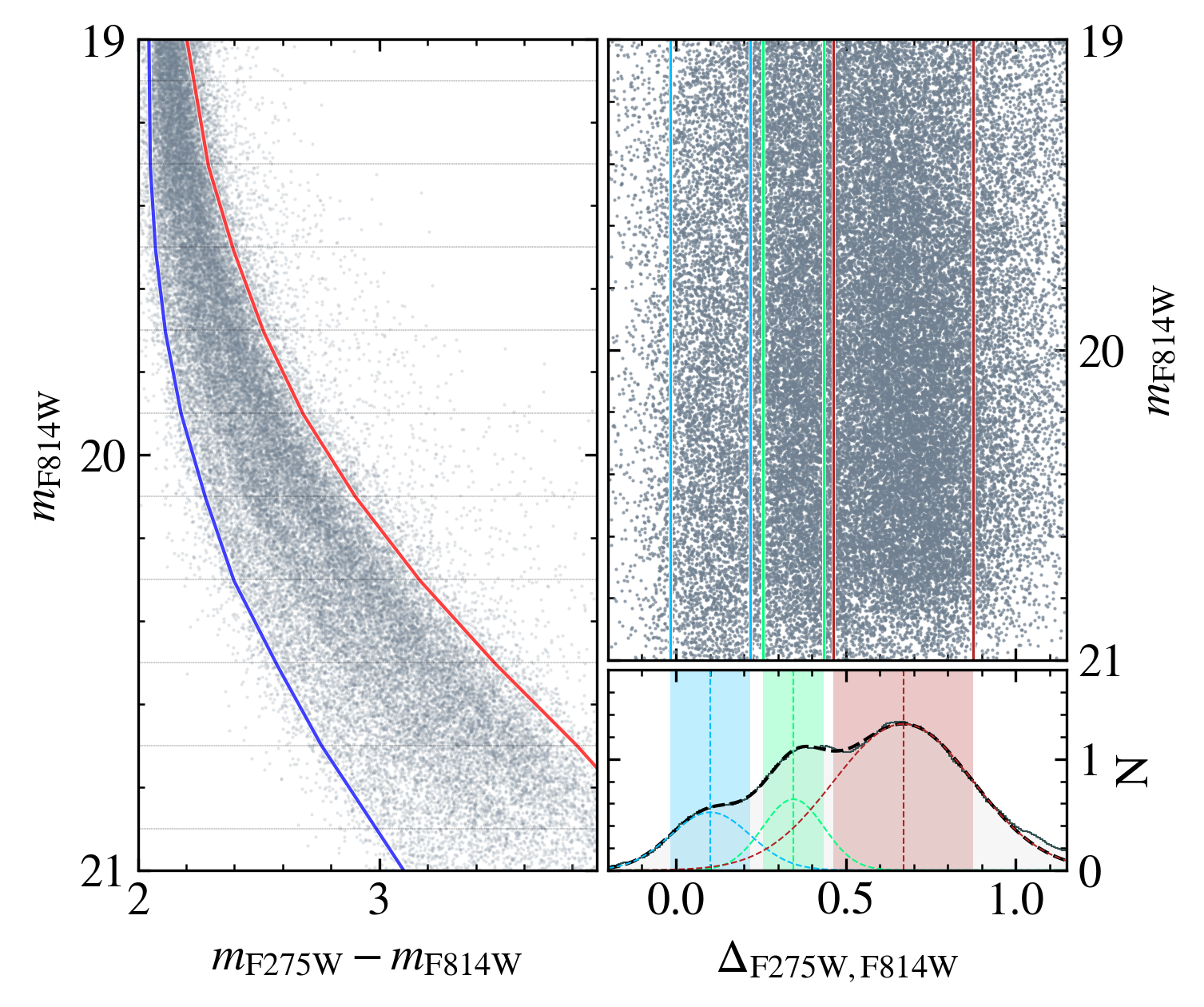}
    \caption{Left: color-magnitude diagram in $m_{\rm F275W}-m_{\rm F814W}$ color of NGC\,2808 below the main-sequence turn-off. The blue and red lines follow the $3^{\rm rd}$ and $97^{\rm th}$ percentiles of the color distribution at a given $m_{\rm F814W}$ magnitude. Top right: rectified color-magnitude diagram. Bottom right: triple Gaussian model fit to the rectified color distribution. Blue, green and red regions represent the $\pm 1 \sigma$ intervals for each Gaussian, and are used to select a sample of stars of each population (bMS, mMS and rMS respectively). See text for details.}
    \label{fig:sel1}
    \includegraphics[width=\columnwidth]{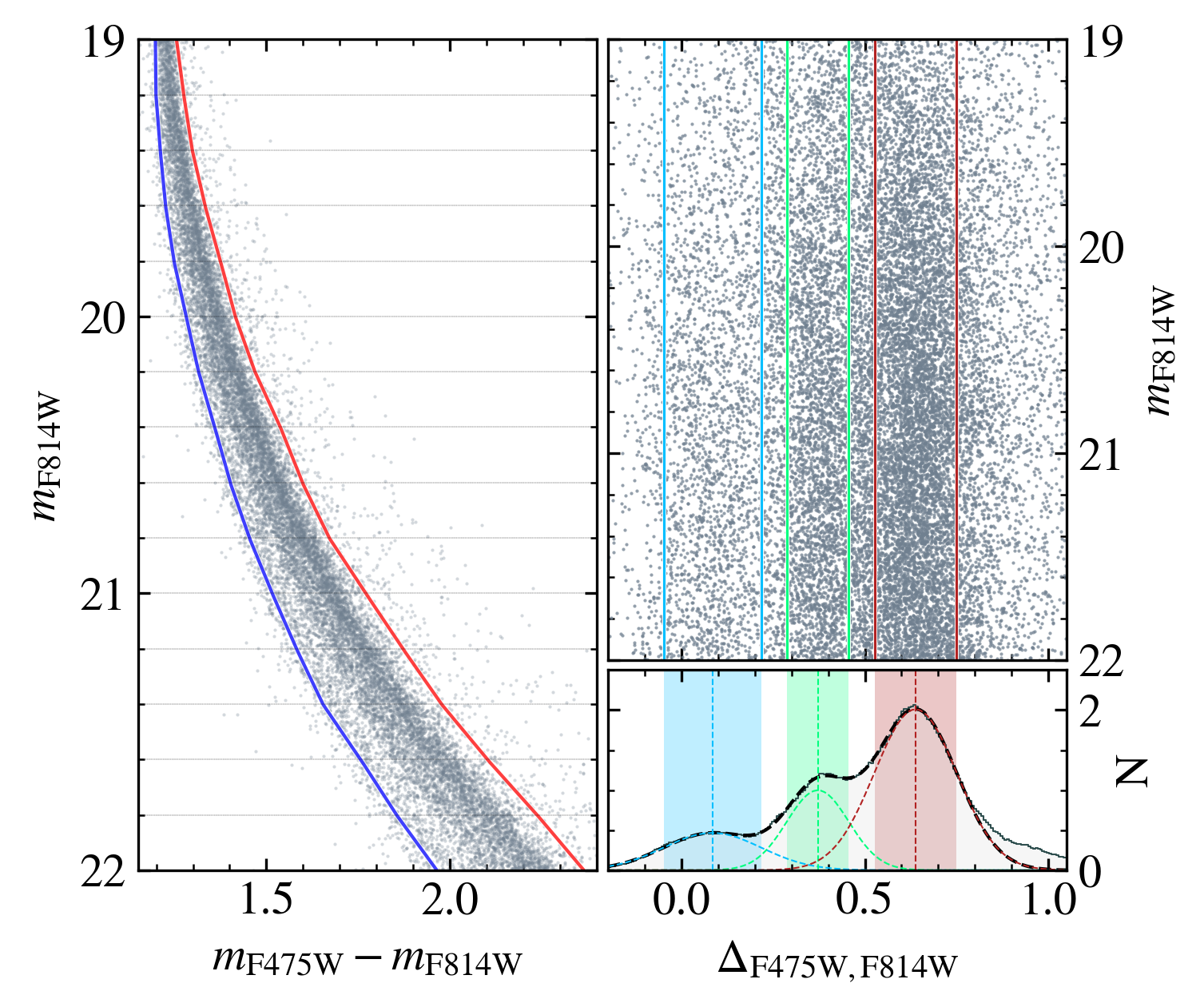}
    \caption{Similar to Fig.\,\ref{fig:sel1}, but for the $m_{\rm F475W}-m_{\rm F814W}$ color-magnitude diagram.}
    \label{fig:sel2}
\end{figure}

The selection of the three samples of stars from each sequence was done separately on two relevant color-magnitude diagrams. The $m_{\rm F275W}-m_{\rm F814W}$ diagram is relative to fields F0 and F1, while the $m_{\rm F475W}-m_{\rm F814W}$ refers to F1 and F2. The color-magnitude diagrams shown in {the} left panels of Figs.\,\ref{fig:sel1} and \ref{fig:sel2} were rectified using the blue and red lines defined as the $3^{\rm rd}$ and $97^{\rm th}$ percentile of the color distribution, calculated in steps of 0.2\,$m_{\rm F814W}$, after a sigma-clipping procedure was applied to remove outliers. The rectified diagrams are shown in the top right panels of each figure. We then fitted a triple Gaussian model on the rectified color distributions ($\Delta_{\rm F275W,F814W},\Delta_{\rm F475W,F814W}$), as shown in the bottom right panels. The shaded blue, green and red regions represent our selection for the bMS, mMS and rMS respectively, and are defined within the $\pm 1 \sigma$ intervals, with $\sigma$ being the standard deviation of each Gaussian.

Note that priority in the population tagging is first given to stars for which we can use the $\Delta_{\rm F475W,F814W}$ pseudo color, since the F475W and F814W filters fully cover the analyzed cluster region beyond $2.5r_{\rm h}$. In addition, images taken with these two filters are deep and allow us to separate the populations along the MS over a wider range of masses than with any other filter combination at disposal. Since stars in field F0 do not have F475W observations, and the available F438W images are too shallow, for them we rely on the $\Delta_{\rm F275W,F814W}$ pseudo color. Due to these
considerations, the tagging of stars for which both selections can be used is based only on the $\Delta_{\rm F475W,F814W}$ pseudo color. Also note that some residual population cross-contamination is unavoidably present in our selections;\ in particular the mMS selection is likely partially contaminated by both rMS and bMS stars, which leads to a lower signal for that group in our kinematic analysis.

\begin{figure}
    \centering
    \includegraphics[width=\columnwidth]{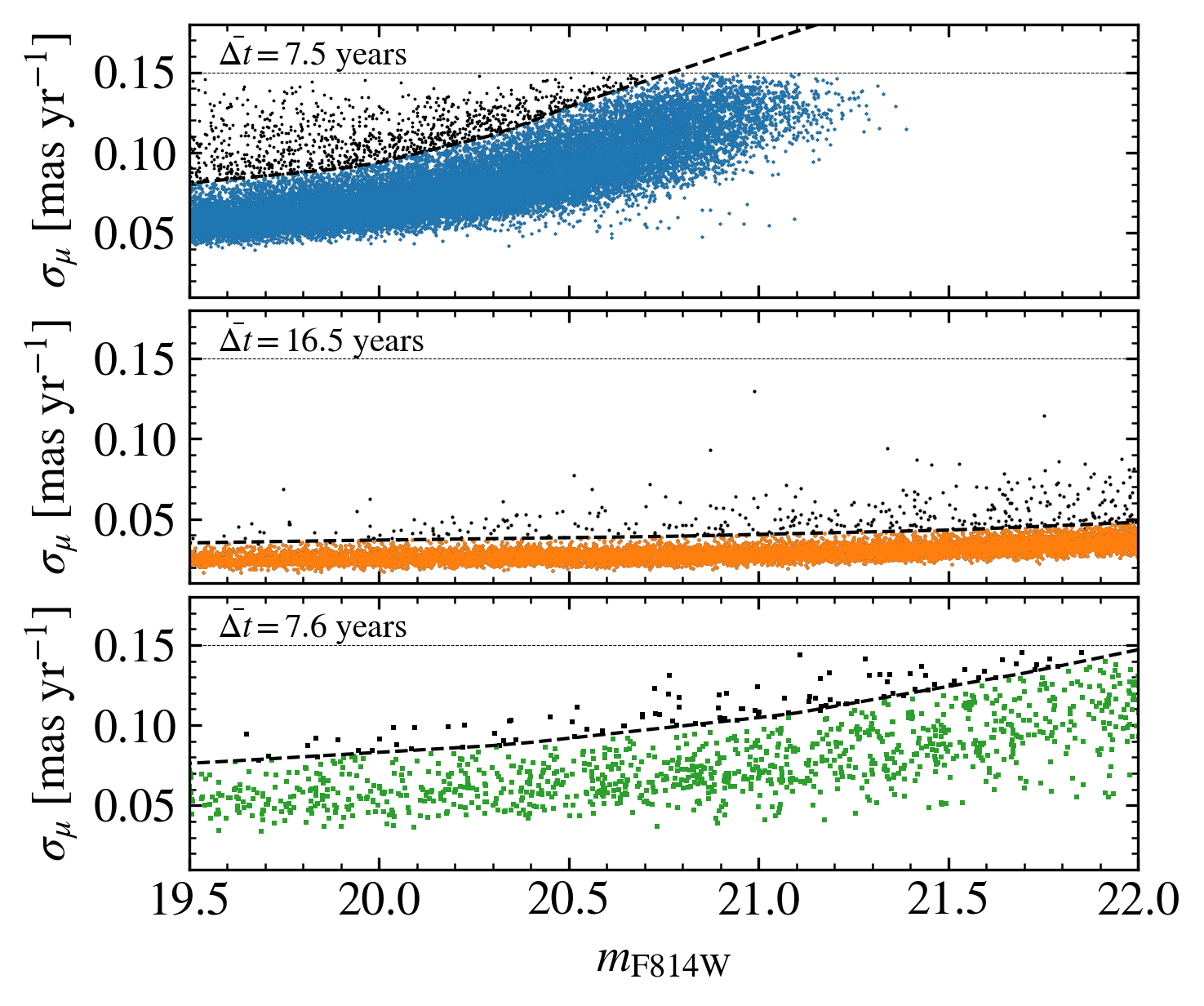}
    \caption{{Combined} proper-motion errors as a function of $m_{\rm F814W}$.
    The three panels show the three different selections (blue, orange and green) relative to the fields F0, F1 and {F2}, respectively.
    The thick black lines represent our cuts. We also excluded stars with errors larger than 0.15\,mas\,yr$^{-1}$ (black, horizontal dashed line). The average {temporal} baseline of sources in each field is reported on the top right corner of each panel. The majority of the sources in F1 have been re{-}observed in GO-15587, extending the baseline to $\sim$\,17\,yr, which allow{s} for a more precise estimate of the proper motions. 
    See text for details.}
    \label{fig:pms_err}
\end{figure}

\begin{figure}
    \centering
    \includegraphics[width=\columnwidth]{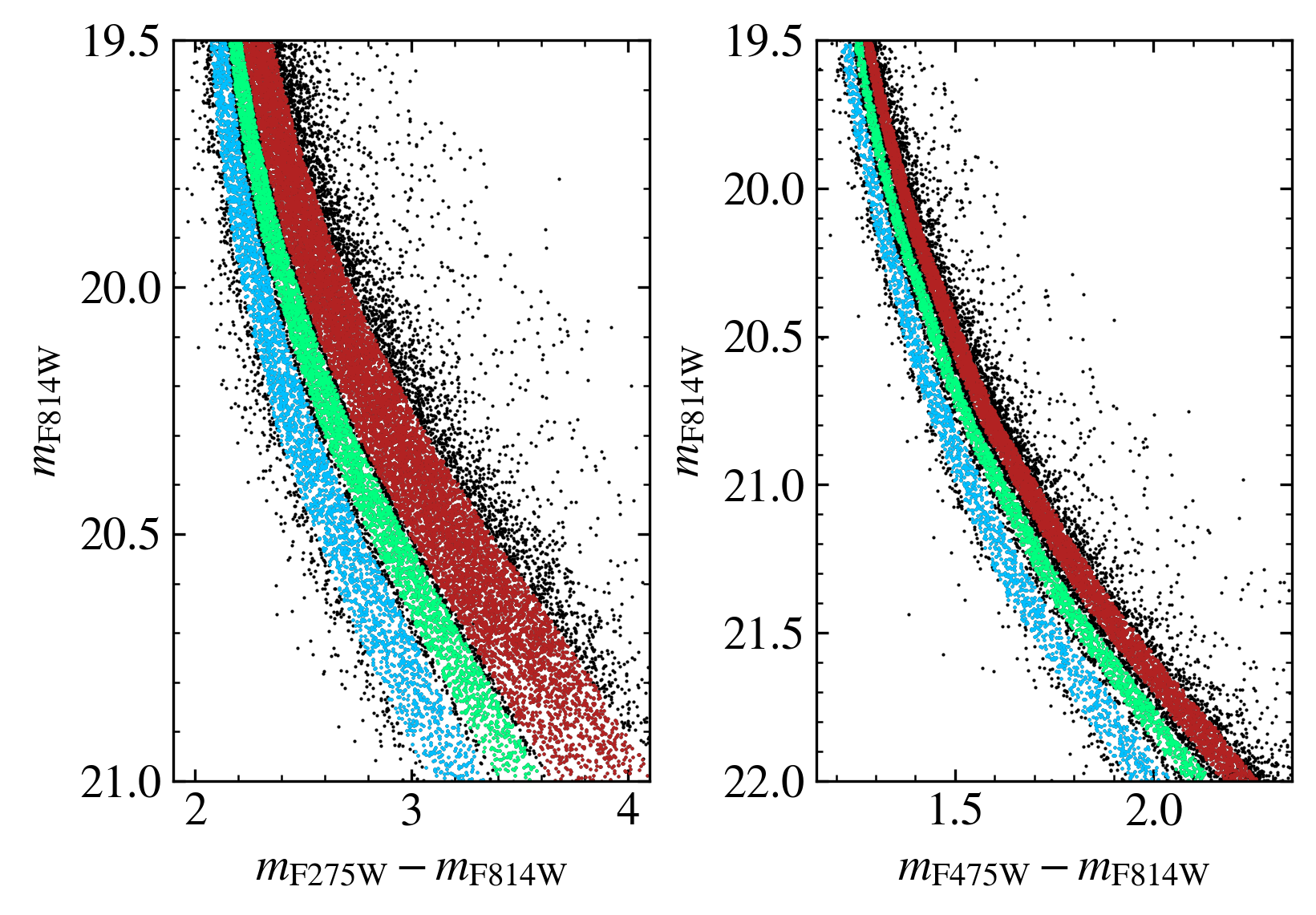}
    \caption{$m_{\rm F275W}-m_{\rm F814W}$ (left) and $m_{\rm F475W}-m_{\rm F814W}$ (right) color-magnitude diagrams of the main sequence of NGC\,2808 with the three samples of stars from each population highlighted in blue, green and red for the bMS, mMS and rMS{,} respectively.}
    \label{fig:cmd_mpop}
\end{figure}

A further selection based on the {combined} proper-motion error has been applied as {shown} in Fig.\,\ref{fig:pms_err}. The errors display different trends depending on the depth of coverage and the total {temporal} baseline available for the proper-motion fit. The three panels show the three different selections (in blue, orange and green) corresponding to the fields F0, F1 and {F2}, respectively. The thick black lines represent our cuts and separate well measured sources (colored) from those that are rejected (black). We also excluded stars with errors larger than 0.15\,mas\,yr$^{-1}$ (black, horizontal dashed line) from all the fields, which is about half of the velocity dispersion in the cluster center. The field are sorted by distance from the center of the cluster, hence the lower source density in field {F2}.  On {the} top of each panel{,} we report the average {temporal} baseline $\Delta t$ for each field.  The smaller errors seen in field F1 are due to the larger {temporal} baseline enabled by the GO-15587 observations. Despite the same {temporal} baseline of about 8\,years, field F0 and {F2} proper-motion error distributions are slightly different, with {F2} having smaller errors at a given magnitude. This is mostly due to the extreme crowding conditions in field F0 that result in higher background noise.

Figure\,\ref{fig:cmd_mpop} shows the two color-magnitude diagrams with the sample of stars from each populations colored in blue, green and red for bMS, mMS, and rMS{,} respectively.

\subsection{Velocity dispersion profiles and energy equipartition}

\begin{figure}
    \centering
    \includegraphics[width=\columnwidth]{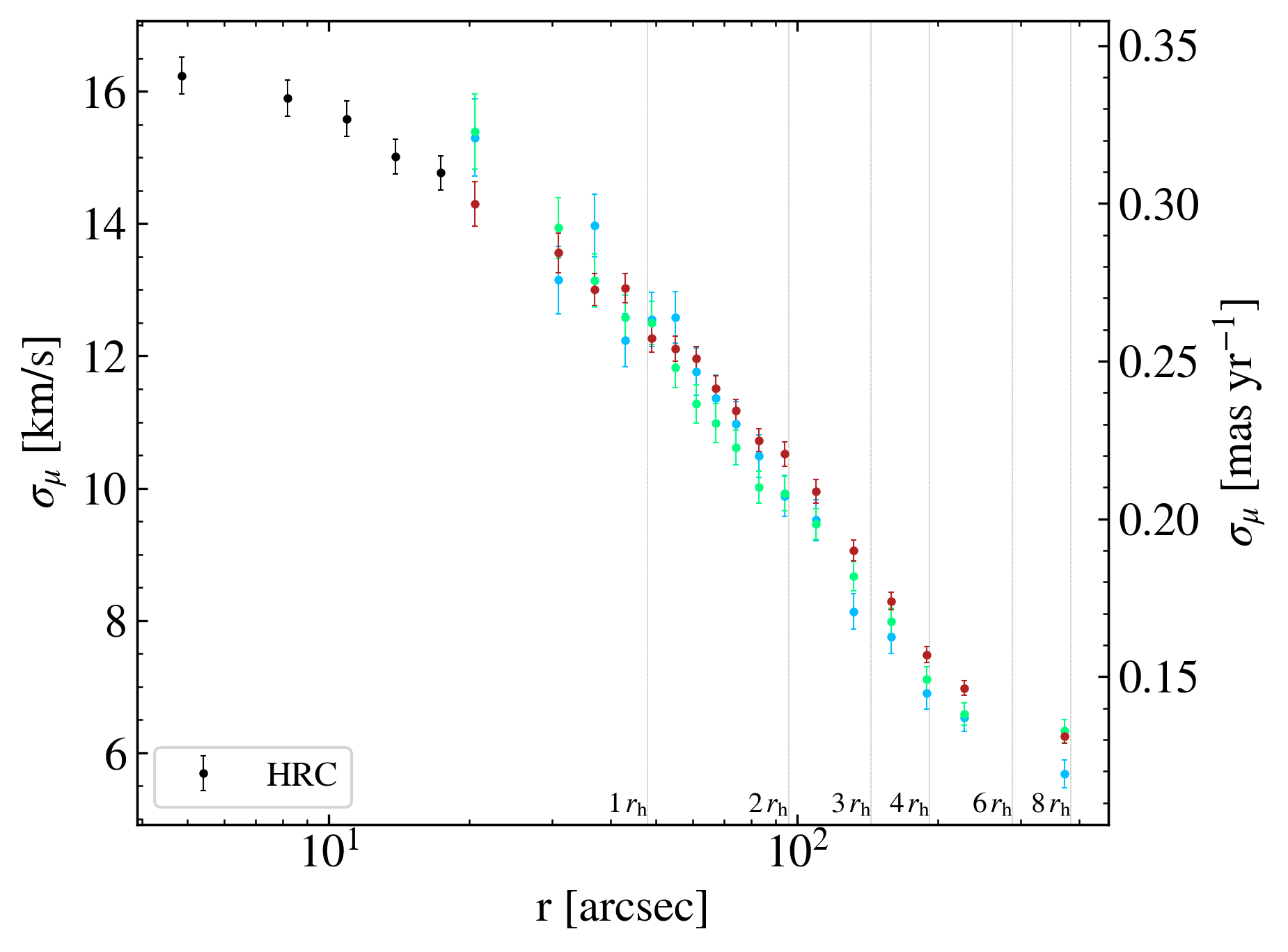}
    \caption{Combined velocity dispersion profiles for the three populations, color{-}coded according to Fig.\,\ref{fig:cmd_mpop}. Black points represent HRC data for which the MPs cannot be distinguished. See text for details.}
    \label{fig:sigma_mu}
\end{figure}

\begin{figure}
    \centering
    \includegraphics[width=\columnwidth]{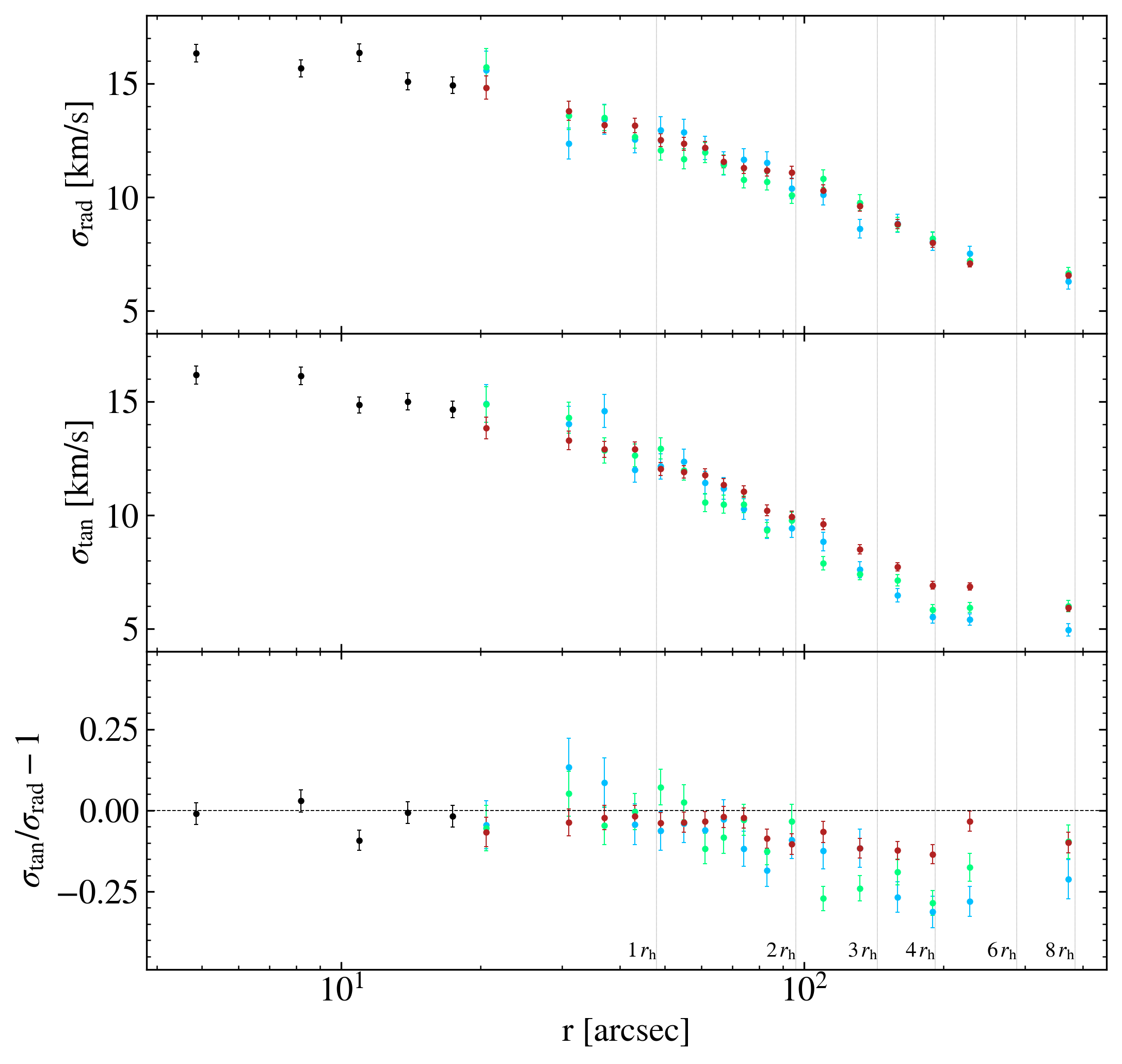}
    \caption{Top and Middle panels: Similar to Fig.\,\ref{fig:sigma_mu}, but for the radial (top) and tangential (middle) components. Bottom: radial profile of the velocity anisotropy (defined as $\sigma_{\rm tan}/\sigma_{\rm rad}-1$). See text for details.}
    \label{fig:sigma_tr}
\end{figure}

\begin{figure}
    \centering
    \includegraphics[width=\columnwidth]{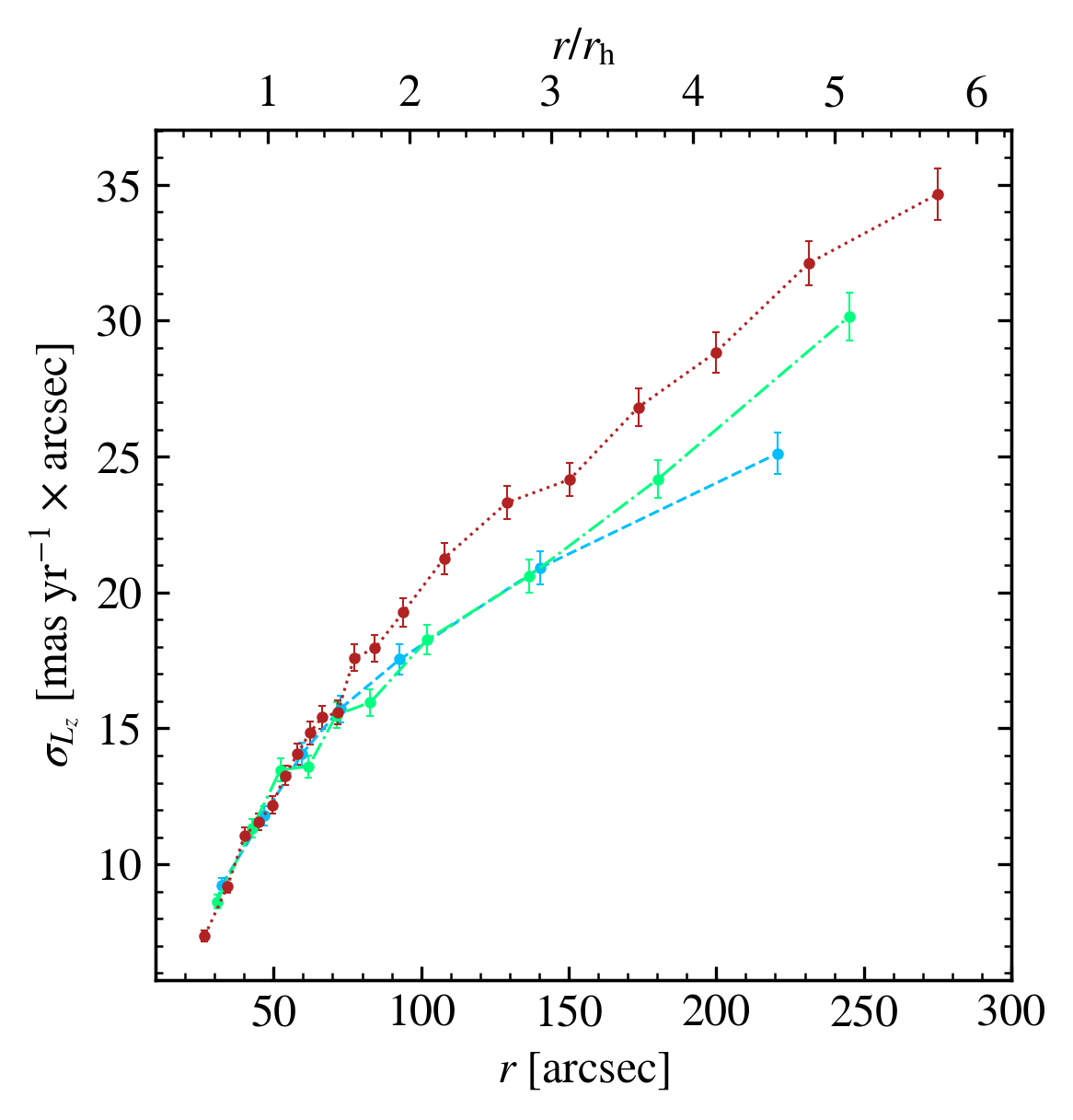}
    \caption{Radial variation of the angular-momentum dispersion for the three populations, with blue, green and red representing the bMS, mMS and rMS, respectively.}
    \label{fig:slz}
\end{figure}

\begin{figure*}
    \centering
    \includegraphics[width=\textwidth]{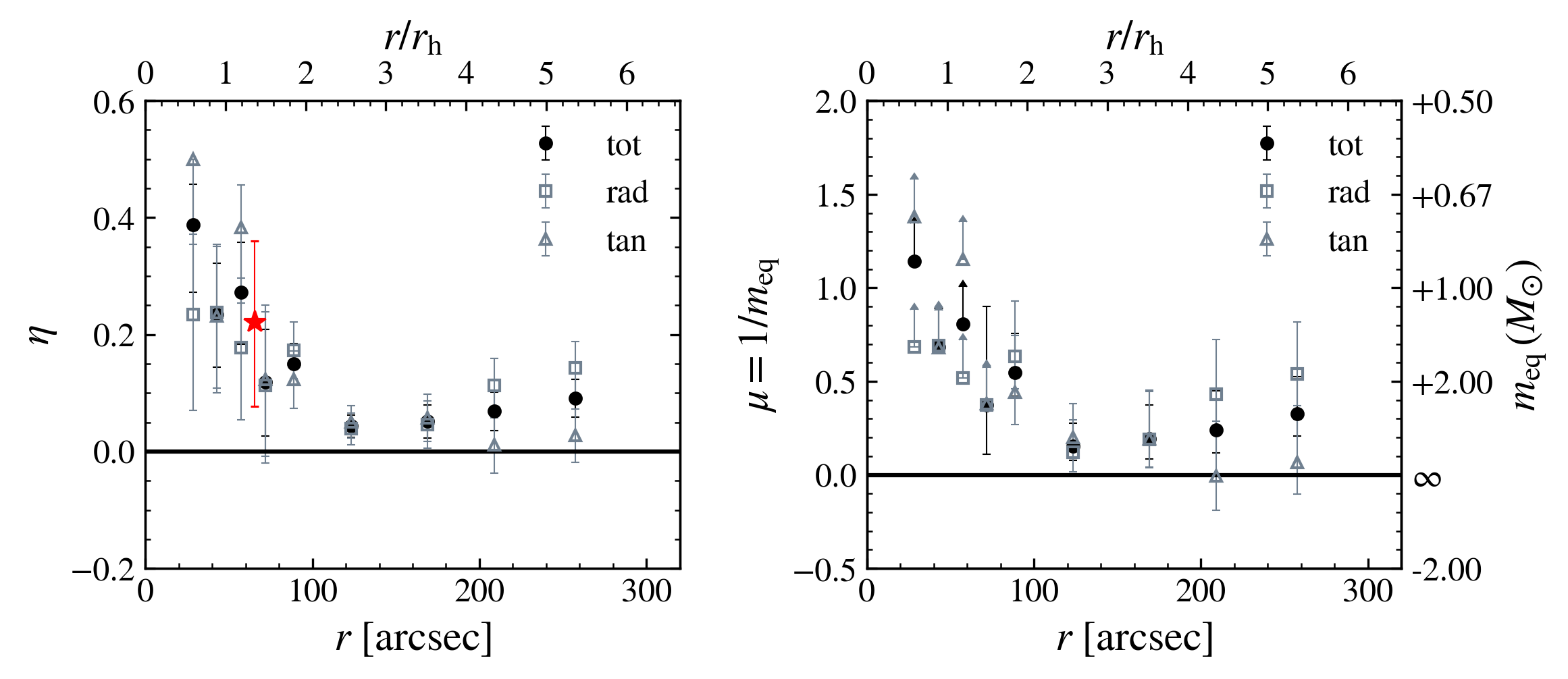}
    \caption{The radial variation of the degree of energy equipartition as measured by the parameter $\eta$ (left panel) and by the equipartition mass parameter, $m_{\rm eq}$ (right panel). Square and triangle markers represent the radial variation of the degree of energy equipartition calculated separately for the radial and the tangential components of the velocity dispersion, respectively. For bins where the mass range does not provide enough coverage, we can only estimate upper limits for $m_{\rm eq}$ (lower limits in $\mu$) and are marked with an arrow. The red star marker in the left panel represents the previous measurement of $\eta$ by \cite{2022ApJ...936..154W}.}
    \label{fig:eq_r}
\end{figure*}

Using only sources that are well measured  both in terms of photometry and astrometry, we computed the intrinsic velocity dispersion profiles for the three populations as in \cite{2022ApJ...934..150L}, i.e. maximizing the following likelihood:
\begin{multline}    
\mathrm{ln}{ \mathcal L } = -\frac{1}{2} \sum_{n} \left[ \frac{{\left({v}_{\mathrm{rad},{\rm{n}}}-{v}_{\mathrm{rad}}\right)}^{2}}{{\sigma }_{\mathrm{rad}}^{2}+{\epsilon }_{\mathrm{rad},{\rm{n}}}^{2}}+\mathrm{ln}\left({\sigma }_{\mathrm{rad}}^{2}+{\epsilon }_{\mathrm{rad},{\rm{n}}}^{2}\right)\right. \\
\left.+\frac{{\left({v}_{\tan,{\rm{n}}}-{v}_{\tan }\right)}^{2}}{{\sigma }_{\tan }^{2}+{\epsilon }_{\tan,{\rm{n}}}^{2}}+\mathrm{ln}\left({\sigma }_{\tan }^{2}+{\epsilon }_{\tan,{\rm{n}}}^{2}\right)\right],
\label{eq:lk}
\end{multline}
where $(v_{\rm rad,n}, v_{\rm tan,n})$ and $(\epsilon_{\rm rad,n}, \epsilon_{\rm tan,n})$ are the radial and tangential components and relative errors of the proper motion of the $n$-th star, and $(v_{\rm rad}, v_{\rm tan})$ and $(\sigma_{\rm rad}, \sigma_{\rm tan})$ are the radial and tangential mean motions and velocity dispersions of the population that we want to estimate. In addition to the radial and tangential 
components, we also estimated the combined velocity dispersion $\sigma_\mu$ by imposing in Eq.\,\ref{eq:lk} $\sigma_{\rm rad} = \sigma_{\rm tan} = \sigma_\mu$.

We computed the velocity dispersions for different radial distances, dividing the data into radial bins such that the number of stars in the less populated sequence (bMS) is always in the range 150--250. The ACS/HRC data were not divided by populations, since the available filters 
do not allow us to separate
the three main-sequences in the color-magnitude diagram.

We used the Markov Chain Monte Carlo (MCMC) method \texttt{emcee} to sample the parameter space and obtain the posterior probability distribution functions (PDFs) for $\sigma_{\rm rad}$, $\sigma_{\rm tan}$ and $\sigma_\mu$.  The MCMC was run with 20 walkers for 5000 steps, and we rejected the first 200 steps.  The medians of the PDFs give the best-fit values, and the errors are given by the average of the $16^{\rm th}$ and $84^{\rm th}$ percentiles of the residuals around the median.  The velocity dispersions were corrected as in \cite{2015ApJ...803...29W} to account for the bias in the maximum-likelihood estimators.

Fig.\,\ref{fig:sigma_mu} shows the combined velocity dispersion profile from about 0.1\,$r_{\rm h}$ out to $\sim$\,8\,$r_{\rm h}$ for the three populations, using the same color scheme as in the previous plots. Black points represent HRC data that include the contribution of all populations. Vertical dashed lines show multiples of the half-light radius of the cluster ($r_{\rm h}=48^{\prime\prime}$). The radial and tangential components of the velocity dispersion, together with their ratio, are shown in Fig.\,\ref{fig:sigma_tr}. The black points are based on HRC data.

In Fig.\,\ref{fig:slz}, we further explore the differences between the anisotropy of 1G and 2G stars by using the radial variation of a new parameter introduced by \cite{2025arXiv250502921A}, $\sigma_{L_z}$, defined as the dispersion of the angular momentum $L_z=v_T \times r$, where $v_T$ is the tangential component of the velocity calculated from the proper-motion measurements and $r$ is the projected distance from the center of the cluster.
The $\sigma_{L_z}$ parameter traces in a more direct way the distribution of tangential velocities in the cluster and enhances differences at large radii that might not be apparent with differences in velocity anisotropy. In particular, we observe small differences in radial velocity dispersion between the populations, and the tangential component drives the differences in velocity anisotropy, the $\sigma_{L_z}$ parameter traces these differences without considering the radial velocity dispersion as in the case of the velocity anisotropy.
As shown in \cite{2025arXiv250502921A}, a larger radial anisotropy is associated with smaller values of $\sigma_{L_z}$. The radial profiles of $\sigma_{L_z}$ for the 1G and the 2G populations show that the two population kinematics differ in the outer regions, with the 2G populations characterized by smaller values of $\sigma_{L_z}$, providing additional evidence of a larger radial anisotropy of the 2G compared to the 1G.\\

The observational data presented in this paper allow us to study the degree of energy equipartition in NGC\,2808. Evolution towards energy equipartition is one of the fundamental dynamical effects of two-body relaxation in star clusters: as a consequence of energy exchange during stellar encounters, clusters gradually evolve from an initial state of no velocity equipartition  (where stars with different masses have all the same velocity dispersion) towards a dynamical configuration where, at a given distance from the cluster's center, low-mass stars are characterized by a larger velocity dispersion than high-mass stars \citep[see, e.g.,][]{1969ApJ...158L.139S,1987degc.book.....S}. 

A number of numerical investigations have been devoted to the study of energy equipartition in globular clusters \citep[see, e.g.,][]{2013MNRAS.435.3272T}, how the evolution towards energy equipartition is affected by various cluster properties such as the initial kinematic properties \citep[see, e.g.,][]{2021MNRAS.504L..12P,2022MNRAS.509.3815P,2022MNRAS.512.2584L,2024A&A...689A.313P} and the fraction of stellar remnants \citep{2023MNRAS.525.3136A}, and the link between the degree of energy equipartition and a cluster's core collapse \citep{2018MNRAS.475L..96B}. More recently. a few studies \citep{2021MNRAS.502.4290V,2024MNRAS.534.2397L} have extended the investigation of energy equipartition to multiple stellar populations and shown that 2G stars in the intermediate/outer cluster's regions may be characterized by a stronger degree of equipartition than 1G stars.

The observational investigation of the state of energy equipartition has become possible mainly thanks to \hst proper-motion measurements of stars spanning a mass range broad enough to allow the study of the dependence of the velocity dispersion on stellar mass 
\citep[see, e.g.,][see also \citealt{2018ApJ...853...86B} for observations consistent with the predictions of numerical simulations]{2018ApJ...861...99L,2019ApJ...873..109L,2022ApJ...936..154W}.

For our analysis of the degree of energy equipartition in NGC\,2808, we have calculated the stellar masses as follows.
To account for the different masses of bMS, mMS and rMS stars at a given magnitude, we interpolated the average mass values in Table~3 of \cite{2012A&A...537A..77M} for the three populations with a quadratic spline within their available range. We also extrapolated the values to cover the full range explored in our analysis that is outside their table. We verified that the derived masses for 1G rMS stars closely match those from the BaSTI isochrone throughout our mass range. Next, we used the relations in Table~6 of \cite{2016MNRAS.463..449S} to  compute the population ratios as a function of distance from the center of the cluster. For each star, we compute the bMS, mMS and rMS masses at the star's corresponding magnitude as predicted by the spline fits. Finally, the mass assigned to a star is obtained as the weighted mean of these three masses, where the weights are given by the population ratios at the distance of the star. The available mass range varies with radius, with the minimum and maximum average masses being 0.23\,M$_\odot$ and 0.78\,M$_\odot$, respectively.

Note that we repeated the same energy equipartition analysis by only using 1G masses for all stars, instead of assigning a population-weighted mean mass to them, and found the results in agreement between each other to better than $1 \sigma$.

In Fig.\,\ref{fig:eq_r}, we show the radial variation of the degree of energy equipartition as measured by the parameter $\eta$ (left panel), defined from the power-law fit of the dependence of the velocity dispersion, $\sigma$, on the stellar mass, $m$, $\sigma \propto m^{-\eta}$ \cite[see, e.g.,][]{2013MNRAS.435.3272T}, and the equipartition mass parameter $m_{\rm eq}$ \citep[introduced by][]{2016MNRAS.458.3644B} and its reciprocal $\mu=1/m_{\rm eq}$, that allow for continuous estimation of the degree of equipartition, from positive to negative slopes in the mass - velocity dispersion relation \citep[right panel, see][]{2023MNRAS.525.3136A}\footnote{Examples of energy-equipartition estimates using these 2 parameters for other GCs can be found in Figs.~7 and 11 of \cite{2022ApJ...936..154W}.}.  These were estimated with proper-motion data of the full sample of main-sequence stars, including those without population classification.

In each radial bin, we follow a discrete fitting approach using the likelihood function from \cite{2022ApJ...936..154W} \citep[see also][]{2023MNRAS.525.3136A} and a uniform prior for $-0.5 < \eta < 0.5$ and $-10<\mu<10$ (for $\mu\geq 10$, we expect that all stars in a radial bin are in full energy equipartition, i.e. $m_{\rm eq}=0.1$). We compare our measured radial variation of $\eta$ with the estimated value from \cite{2022ApJ...936..154W} (red star). This value comes from all stars within their sample, and it is consistent with our measurements. For the three innermost radial bins, the lower mass limit is $\sim$\,$0.57$\,M$_{\odot}$. The resulting mass range limits our constraints in the $\mu$ parameter, as indicated with lower limits in the right panel of Fig.\,\ref{fig:eq_r}. For radial bins beyond $1.5\,r_{h}$, our sample has a wide enough mass range to constrain the values of $\mu$.

Both panels also show the radial variation of the degree of energy equipartition calculated separately for the radial and the tangential component of the velocity dispersion. A similar study of the radial variation of the level of equipartition for individual stellar generations is not possible with our data because different generations can be identified along a very narrow range of stellar masses.

\section{Discussion}
\label{sec:th}

The velocity dispersion and anisotropy radial profiles shown in Figs.\,\ref{fig:sigma_mu}, \ref{fig:sigma_tr}, and \ref{fig:slz} show clear evidence of different kinematical behaviors among the three populations. In particular, in the outer regions of the cluster, the two 2G populations (bMS and mMS) are characterized by a higher velocity anisotropy than the 1G (rMS) population. The differences in the velocity anisotropy are due to the smaller tangential velocity dispersion of the 2G populations. Both of these differences are consistent with the predictions of numerical simulations that follow the dynamical evolution of multiple-population GCs. Specifically, simulations starting with 2G stars being initially more centrally concentrated in the inner regions of a 1G system \citep[see, e.g.,][]{2015ApJ...810L..13B,2021MNRAS.502.4290V,2025arXiv250502921A} show that, as the 2G stars diffuse outwards, they acquire a radially anisotropic velocity distribution, while the 1G population is characterized by an isotropic or slightly anisotropic velocity distribution. Simulations also predict that these differences are due to the smaller tangential velocity dispersion of 2G stars while no significant differences are found in the radial velocity dispersions of 1G and 2G stars. Kinematic differences consistent with these theoretical predictions have also been found in a few other observational studies on the subject \citep[see, e.g.,][]{2015ApJ...810L..13B,2023ApJ...944...58L,2024A&A...685A.158C,Dalessandro_2024,2025MNRAS.537.2342C}.

As for energy equipartition, both panels of Fig.\,\ref{fig:eq_r} show a similar radial variation in the degree of energy equipartition, indicating a dynamical state closer to full energy equipartition in the inner regions and less energy equipartition at increasing distances from the center of the cluster. The observed radial profile is consistent with the expected variation due to the relaxation timescale increasing with the distance from the center \citep[see also][]{2022ApJ...936..154W,2018ApJ...861...99L}. 

A recent study by \cite{2023MNRAS.525.3136A} explored the connection between the degree of equipartition in the clusters' inner regions and the fraction of the clusters' mass in stellar mass black holes (BHs). In this case, the relatively strong degree of energy equipartition in the inner regions of NGC\,2808 would suggest a small fraction of BHs, consistent with the findings of \cite{2024MNRAS.529..331D} who estimated the mass fraction of BHs in NGC\,2808 to be about 0.08 per cent. 

No differences between the degree of equipartition in the radial and tangential components are found within $r/r_{\rm h} \lesssim 4$; in the outermost regions ($r/r_{\rm h} >4$) data hint at a possible difference between the two measures of equipartition, with the radial component possibly being characterized by a slightly stronger degree of equipartition than the tangential component.

 {We find} no systematic difference between the anisotropy of the two 2G populations {within $\sim 3 r_{\rm h}$}; {outside} $3r_{\rm h}$, data hint at a possible trend for the 2G bMS radial anisotropy to be slightly greater than that of the 2G mMS, but the differences do not appear statistically significant.

\section{Conclusions}
\label{sec:con}
 
This paper presents a study of the internal kinematics of the globular cluster NGC\,2808
based on multi-epoch \hst observations spanning a total {temporal} baseline of 17 years. The data used in this analysis cover a region from the cluster core out to about {8}\,$r_{\rm h}$---about 5 times the spatial coverage of previous studies. Thanks to state-of-the-art techniques, we were able to measure high-precision photometry and astrometry for stars down to $0.25$\,M$_\odot$.
The broad range of stellar masses for which proper-motion measurements are possible allowed us to measure the degree of energy equipartition and its variation with distance from the cluster's center. We find that the degree of energy equipartion decreases at larger distances from the cluster's center, in agreement with  theoretical expectations, due to the local relaxation time increasing with clustercentric distance. In the cluster's outer regions, we find hints of a difference in the degree of equipartition between the radial and tangential components of the velocity dispersion \citep[see][for numerical studies predicting differences between the tangential and radial equipartitions in the outer regions of the cluster; see also \citealt{2015ApJ...810L..13B, 2018ApJ...853...86B,   2018ApJ...861...99L} for the first observational studies]{2021MNRAS.504L..12P,2022MNRAS.509.3815P}.

Color-magnitude diagrams in the $m_{\rm F475W}-m_{\rm F814W}$ and $m_{\rm F275W}-m_{\rm F814W}$ colors show that the main sequence can be split into three components, a 1G population and two 2G populations with different He content. We measured the velocity dispersion of these three subpopulations as a function of the distance from the cluster center, and we found that the 2G populations are radially anisotropic due to a smaller tangential velocity component, confirming the previous findings of \cite{2015ApJ...810L..13B} at $r \lesssim 2 r_{\rm h}$. We traced the velocity dispersion profiles out to $\sim$\,8\,$r_{\rm h}$, showing that the anisotropy is larger in the external regions ($r>2\,r_{\rm h}$).
Differences between the anisotropy of 1G and 2G stars were also revealed by measurements of the dispersion in the angular momentum in the plane of the sky, $\sigma_{Lz}$, a new parameter recently introduced by \cite{2025arXiv250502921A}.

Our findings are in agreement with the predictions of numerical simulations \citep[see, e.g.][]{2015ApJ...810L..13B,2021MNRAS.502.4290V} of the dynamical evolution of MPs in GCs, which suggest that 2G stars, initially more concentrated in the inner regions, gradually diffuse outward and develop a radially anisotropic velocity distribution. In contrast, the 1G population maintains an isotropic or weakly anisotropic distribution.  Simulations also predict that these kinematical differences arise from a smaller tangential velocity dispersion of the 2G populations \citep[see, e.g.][]{2015ApJ...810L..13B,2021MNRAS.502.4290V,2024MNRAS.534.2397L}, while no significant discrepancies are predicted in the 
radial component of the velocity dispersions between 1G and 2G stars, which is in agreement with our observations.  These differences are expected to be present in particular in relatively dynamical{ly} young clusters like NGC\,2808, in which observational studies \citep[see, e.g.][]{2016MNRAS.463..449S,2019ApJ...884L..24D,2023MNRAS.520.1456L,Dalessandro_2024} have found the 2G to be still spatially more centrally concentrated than the 1G.

The extension of the study of energy equipartition to the individual stellar generations would provide new insights into the dynamics of multiple populations \citep[see][]{2021MNRAS.502.4290V,2024MNRAS.534.2397L}; however, this is not feasible with the data used in this paper, since the identification of the three stellar generations is only possible for stars in a narrow mass range, thus preventing a reliable estimate of the degree of energy equipartition.
\\

While our analysis shows differences between the 1G and the 2G kinematics, the current observational data do not provide enough evidence to draw definitive conclusions regarding potential differences between the kinematics of the two 2G subgroups (the mMS and the bMS). 
An \hst study of the spatial distribution of the mMS and the bMS by \cite{2016MNRAS.463..449S} has shown that the bMS is more centrally concentrated than the mMS (and both the 2G bMS and the mMS are more concentrated than the 1G rMS); a stronger radial anisotropy of the bMS with respect to the mMS, if confirmed, could be linked to initial differences in the spatial concentration of the bMS and the mMS subpopulations, consistent with the trend observed by \cite{2016MNRAS.463..449S} (see also \citealt{2019MNRAS.489.3269C} for hydro/N-body simulations of multiple stellar population formation finding a similar trend). Evidence of dynamical differences between 2G subgroups could thus provide a new constraint for models of the formation and evolution of multiple-population globular clusters. Further observational investigations of this aspect are necessary before drawing any conclusions about possible differences between the 2G populations and the implications for models of the formation of multiple populations.

We make our astro-photometric catalogs publicly available to the community on VizieR\footnote{\url{https://vizier.cds.unistra.fr/viz-bin/VizieR}.}.

\begin{acknowledgments}
Based on observations with the NASA/ESA {\it Hubble Space Telescope\/} obtained from the Data Archive at the Space Telescope Science Institute (STScI), which is operated by the Association of Universities for Research in Astronomy, Incorporated, under NASA contract NAS5-26555.
This work presents results from the European Space Agency (ESA) space mission \gaia. \gaia data are being processed by the \gaia Data Processing and Analysis Consortium (DPAC). Funding for the DPAC is provided by national institutions, in particular the institutions participating in the \gaia MultiLateral Agreement (MLA). The \gaia mission website is https://www.cosmos.esa.int/gaia. The \gaia archive website is https://archives.esac.esa.int/gaia.
Support for Program number GO-15857 and AR-16157 was provided through grants from STScI under NASA contract NAS5-26555.
This project is part of the HSTPROMO (High-resolution Space Telescope PROper MOtion) Collaboration (\url{https://www.stsci.edu/~marel/hstpromo.html}), a set of projects aimed at improving our dynamical understanding of stars, clusters, and galaxies in the nearby Universe through measurement and interpretation of proper motions from \hst, \gaia, and \textit{JSWT}. We thank the collaboration members for the sharing of their ideas and software.
This work is part of the project Cosmic-Lab at the Physics and Astronomy Department ``A. Righi'' of the Bologna University (\url{http://www.cosmic-lab.eu/Cosmic-Lab/Home.html}).
\end{acknowledgments}

\bibliography{bibliography}{}
\bibliographystyle{aasjournal}



\end{document}